\documentclass[preprint,showpacs,aps]{revtex4}
\begin{document}
\draft \title{Excited states of quantum many-body interacting systems: A
variational coupled-cluster description} 
\author{Y. Xian}
\address{School of Physics and Astronomy, The University of
Manchester, Manchester M13 9PL, UK} \date{\today } 

\begin{abstract}

We extend recently proposed variational coupled-cluster method to describe
excitation states of quantum many-body interacting systems. We discuss,
in general terms, both quasiparticle excitations and quasiparticle-density-wave
excitations (collective modes). In application to quantum antiferromagnets, 
we reproduce the well-known spin-wave excitations, i.e. quasiparticle
magnons of spin $\pm 1$. In addition, we obtain new, spin-zero 
magnon-density-wave excitations which has been missing in Anserson's spin-wave 
theory. Implications of these new collective modes are discussed.

\end{abstract}


\maketitle

\section{introduction}

In the preceding papers (Ref. 1 and 2, hereafter referred to as
paper I and II), we proposed a general variational theory for ground
states of quantum many-body interacting systems. Our analysis
extends the well-established coupled-cluster
method (CCM) \cite{hhc,ck,ciz} to a variational formalism in which
bra and ket states are now hermitian to one another, contrast to the
traditional CCM where they are not \cite{arp}. Ever since the CCM was
first proposed, attempts have been made to extend it to a
standard variational formalism, for examples, in the seventies in 
nuclear physics \cite{ps} and later in quantum chemistry \cite{kv}. It 
is perhaps fair to say that progress of this variational approach is 
slow, particularly when comparing with a plethora of
applications made by the traditional CCM over the last 35 years \cite{bart}.
Main difficulties in this variational approach
include {\it ad hoc} approximation truncations and slow convergent
numerical results. In I and II, we provided a new systematic scheme
to overcome these difficulties. In particular, we introduced two 
sets of important {\it bare} distribution 
functions and derived self-consistency equations for these functions;
calculations of physical quantities can all be done in terms of these
functions. This strategy is similar to that employed by another
well-established variational theory, the method of correlated basis
functions (CBF) \cite{fee}, where density distribution functions are
key ingredients. We showed that the traditional CCM is a simple
{\it linear} approximation to one set of bare distribution functions.
We introduced diagrammatic techniques to calculate those distribution
functions to high orders for achieving convergent results;
resummations of infinite (reducible) diagrams 
can now be done by a practical, self-consistent technique. 
Furthermore, in our diagrammatic approach, a close relation
with the CBF method was established and exploited; a possible combination
of these two methods was also proposed. We demonstrated the efficacy
of our variational method by applying to quantum antiferromagnets.
The ground-state properties  of spin-wave theory (SWT) \cite{and} was
reproduced in a simple approximation. Approximation beyond SWT 
by including higher-order, infinite sets of reducible diagrams produced
convergent, improved numerical results for square and cubic lattices and, 
interestingly, it also cures the divergence by SWT in one dimensional system. 

In this article we extend our variational CCM to describe excitation states.
A brief report of some preliminary results has been published \cite{yx3}. 
We investigate two different types of excitation states 
using two approaches. In the first approach, 
we follow the traditional CCM \cite{emr,bpx} to investigate quasiparticle 
excitations, but keeping our ket and bra excited states hermitian to one another.
We then investigate collective modes by adapting Feynman's excitation theory
of phonon-roton spectrum of helium liquid \cite{fey} to our method. In application 
to antiferromagnets, we find that quasiparticle excitations correspond to Anderson's
spin-wave excitations which are often referred to as magnons with 
spin $+1$ or $-1$ \cite{and}. We find collective modes in these
quantum antiferromagnets as longitudinal, spin-zero magnon-density-wave 
excitations which have been missing in Anderson's theory. In our approximation, 
energy spectra of these spin-zero excitations show a large gap for a
cubic lattice (3D) and are gapless in a square lattice (2D). These spectra are
similar to those of charge-density-wave excitations (plasmons) in
quantum plasmas such as electron gases at low temperature \cite{pine}.
More discussion on these collective modes will be given in the final 
section of this article.

\section{ground states by variational coupled-cluster method}

We briefly summarize in this section our variational approach for the 
ground state of a many-body interacting system. Details can be found in I and II.
We take a spin-$s$ antiferromagnetic Heisenberg model
on a bipartite lattice as our model system. The Hamiltonian is given by
\begin{equation}
H=\frac 12\sum_{l,n}{\bf s_l}\cdot {\bf s_{l+n}}\;,
\end{equation}
where index $l$ runs over all lattice sites, $n$ runs over all $z$ nearest-neighbor
sites. We use Coester representation for both ket and bra ground-states, and
write
\begin{equation}
|\Psi_g\rangle=e^S|\Phi\rangle,\quad S=\sum_I F_I C^\dag_I\;;\quad
\langle\tilde\Psi_g|=\langle\Phi|e^{\tilde S},\quad
 \tilde S=\sum_I \tilde F_I C_I,
\end{equation}
where model state $|\Phi\rangle$ is given by the 
classical N\'eel state, $C_I^\dagger$ and $C_I$ with nominal 
index $I$ are the so-called configurational creation and 
destruction operators and are given by, for the spin lattice of Eq.~(1),
\begin{equation}
\sum_I F_I C^\dag_I= \sum_{k=1}^{N/2}
\sum_{i_1...,j_1...}f_{i_1...,j_1...}
\frac{s_{i_1}^{-}...s_{i_k}^{-}s_{j_1}^{+}...s_{j_k}^{+}}{(2s)^k},
\end{equation}
for the ket state. The bra state
operators are given by the corresponding hermitian conjugate of
Eq.~(3), using notation $\tilde F_I = \tilde f_{i_1...,j_1...}$ for
the bra-state coefficients. As before, we have used index $i$ exclusively
for the spin-up sublattice of the N\'eel state and index $j$ for
the spin-down sublattice. The coefficients $\{F_I,\tilde
F_I\}$ are then determined by the standard variational equations as
\begin{equation}
\frac{\delta\langle H\rangle}{\delta \tilde F_I} =
\frac{\delta\langle H\rangle}{\delta F_I} = 0\;,
\quad
  \langle H\rangle \equiv 
  \frac{\langle\tilde\Psi_g| H|\Psi_g\rangle}
       {\langle\tilde\Psi_g|\Psi_g\rangle}\;.
\end{equation}
The important bare distribution functions,
$g_I\equiv\langle C_I\rangle$ and $\tilde g_I\equiv\langle C^\dagger_I\rangle$,
can be expressed in self-consistency equations as
\begin{equation}
g_I = G(\tilde g_J,F_J)\;,\quad \tilde g_I = G(g_J, \tilde F_J)\;,
\end{equation}
where $G$ is a function containing up to linear terms in $\tilde g_J$
(or $g_J$) and finite order terms in $F_J$ (or $\tilde F_J$). 
The Hamiltonian expectation $\langle H\rangle$ of Eq.~(4) can be
expressed as, in general,
a function containing up to linear terms in $g_I$ and $\tilde g_I$ and
finite order polynomial in $F_I$ (or in $\tilde F_I$),
\begin{equation}
\langle H\rangle = {\cal H}(g_I,\tilde g_I, F_I) = 
{\cal H}(\tilde g_I,g_I, \tilde F_I)\;.
\end{equation}
In I and II, as a demonstration, we considered a simple truncation 
approximation in which the correlation operators $S$ and $\tilde S$ 
of Eqs.~(2) and (3) retain only the two-spin-flip operators as
\begin{equation}
S\approx \sum_{i,j}f_{ij}C^\dag_{ij}
 =\sum_{i,j}f_{ij}\frac{s_i^{-}s_j^{+}}{2s}\;,\quad 
\tilde S\approx \sum_{i,j}\tilde f_{ij}C_{ij}
 =\sum_{i,j}\tilde f_{ij}\frac{s_i^{+}s_j^{-}}{2s}\;. 
\end{equation}
The spontaneous magnetization (order parameter) in this two-spin-flip
approximation is given by the one-body density function
$\rho_{ij}$ as
\begin{equation}
\langle s^z_i\rangle =s-\rho\;,\quad \rho = \sum_j\rho_{ij}
 = \sum_j f_{ij}\tilde g_{ij}\;,
\end{equation}
where we have taken the advantage of translational invariance of the
lattice system. For $j$-sublattice, $\langle s^z_j\rangle=\rho -s$.
Within this approximation the SWT result for the correlation 
coefficient can be derived from Eq.~(4) as
\begin{equation}
f_q=\tilde f_q = \frac{1}{\gamma_q}\left[\sqrt{1-(\gamma_q)^2}
-1\right]\;,\quad
\gamma_q = \frac1z\sum_n e^{i\bf{q\cdot r_n}}\;,
\end{equation}
where $f_q$ is the sublattice Fourier transformation of $f_{ij}$ with
$\bf q$ restricted to the magnetic zone, $z$ is the coordination
number of the lattice, and $n$ is the nearest-neighbor index.
Fourier component of the one-body bare distribution function is derived as,
\begin{equation}
\tilde g_q = \frac{\tilde f_q}{1-\tilde f_qf_q}\;.
\end{equation}
Finally, the two-body distribution functions is approximated by,
in the same order,
\begin{equation}
\tilde g_{ij,i'j'}\approx \tilde g_{ij}\tilde g_{i'j'}
  +\tilde g_{ij'}\tilde g_{i'j}\;.
\end{equation}
Approximation beond these SWT formulas produced improved results and were 
given in details in II. For simplicity of our first attempt to discuss 
excitation states, we shall restrict ourselves to these approximations of 
Eqs.~(7-11) in the following.

\section{quasiparticle excitations}

As mentioned in Sec.~I, inspired by the close relation between our
approach and the CBF method, we can investigate quasiparticle-density-wave
excitations by adapting Feynman's excitation theory, as well as usual quasiparticle
excitations by similar approach as in the traditional CCM. One well-known example 
of a quantum system exhibiting similar two kind of excitations is quantum electron
gases \cite{pine}, where quasiparticle excitations are electron
or hole excitations and collective modes are plasmon excitations representing
longitudinal, charge-neutral density fluctuations of those quasielectron and 
holes. In this section we focus on quasiparticle excitations and leave 
discussion of collective modes in the
next section. We will first discuss these excitations in a general
term and then apply to the spin-lattice model of Eq.~(1) as a demonstration.

Following Emrich in the traditional CCM \cite{emr,bpx}, we express
excitation ket-state $|\Psi_e\rangle$ by a linear operator $X$ constructed
from creation operators acting onto the ground state $|\Psi_g\rangle$ as
\begin{equation}
|\Psi_e\rangle = X|\Psi_g\rangle = Xe^S |\Phi\rangle\;,\quad
X=\sum_L x_LC^\dagger_L\;,
\end{equation}
and, unlike the traditional CCM, our bra excitation state is the
corresponding hermitian conjugate, involving only destruction
operators as
\begin{equation}
\langle\tilde\Psi_e| = \langle\tilde \Psi_g|\tilde X =
 \langle\Phi|e^{\tilde S}\tilde X\;,\quad \tilde X=\sum_L\tilde x_LC_L\;.
\end{equation}
In Eqs.~(12) and (13), the ground-state operators $S$ and $\tilde S$ are
as given by Eqs.~(2), $x_L$ and its hermitian conjugate $\tilde x_L$ 
are excitation coefficients. For quasiparticle creation and
destruction operators $C_L^\dagger$ and $C_L$ of Eqs.~(12-13),
we use index $L$ to mark the following important
difference to the ground-state counterparts $C^\dagger_I$ and
$C_I$ of Eqs.~(2). Due to symmetry consideration, some
configuration operators are not included in the correlation operator
$S$ and $\tilde S$ of the ground states but they are
important in the excited states. In our spin lattice example
of Eq.~(1), the ground-state operators of Eq.~(3) always contain even
number of spin-flip operators (each spin-flip-up operator for the
$i$-sublattice always pairs up with one spin-flip-down operator
for the $j$-sublattice) to ensure the total $z$-component
of angular momentum $s^z_{\rm total}=0$. For the excitation
operators, however, the constraints are different.
The single spin-flip operator $s^-_i$ for the
$i$-sublattice (or $s^+_j$ for the $j$-sublattice) will be the
important first term in Eq.~(12) to be discussed in the followings;
the corresponding excitation state $|\Psi_e\rangle$ is in
the $s^z_{\rm total}=-1$ sector (or $+1$ if $s^+_j$ is
used). Therefore, these excitations are referred to as
quasiparticles carrying spin $\pm1$. For our spin lattice models,
we expect that these quasiparticles are the well-known magnons of
spin-wave excitations \cite{and}.

If the ground state $|\Psi_g\rangle$ is exact with energy $E_0$,
the energy difference between excitation state of Eqs.~(12) and
the ground state can be written as,
\begin{equation}
\epsilon = \frac{\langle\tilde\Psi_g|\tilde XHX|\Psi_g\rangle}
     {\langle\tilde\Psi_e|\Psi_e\rangle} -E_0
  =\frac{\langle\tilde\Psi_g|\tilde X [H, X]|\Psi_g\rangle}
     {\langle\tilde\Psi_e|\Psi_e\rangle}\;,
\end{equation}
which involves a commutation. In general, $|\Psi_g\rangle$
is not exact but calculated by approximations.
For our variational ground states of Eqs.~(2), Eq.~(14) can be shown to
remain valid after replacing the exact energy $E_0$ by the variational energy $E_g$
which obeys the following optimal conditions,
\begin{equation}
E_g =\langle H\rangle= \frac{\langle HC^\dagger_I\rangle}{\tilde g_I} =
      \frac{\langle C_IH\rangle}{g_I}\;,
\end{equation}
derived from Eqs.~(4).

To prove Eq.~(14) after replacing the exact $E_0$ by the variational
$E_g$, we first express the 
normalization of excited states of Eqs.~(12) and (13) as an expectation 
value in the ground-states of Eqs.~(2) as,
\begin{equation}
I_e = \langle\tilde\Psi_e|\Psi_e\rangle =I_g \langle \tilde X X\rangle
 = I_g\sum_{L,L'}\tilde x_{L'}x_L \langle C_{L'}C^\dagger_L\rangle\;,
\end{equation}
where $I_g=\langle\tilde \Psi_g|\Psi_g\rangle$. We now consider
a general linear operator $O=O(C^\dagger_I,C_I)$ (a polynomial
of $C^\dagger_I$ and/or $C_I$),
and write
\begin{equation}
O|\Psi_g\rangle = Oe^S|\Phi\rangle = e^S\bar O|\Phi\rangle\;,
\end{equation}
where the similarity-transformed operator,
$\bar O\equiv e^{-S}Oe^S=O(\bar C^\dagger_I,\bar C_I)$,
$\bar C^\dagger_I=C^\dagger_I$ and
\begin{equation}
\bar C_I = e^{-S}Oe^S = C_I+[C_I,S]+\frac{1}{2!}[[C_I,S],S]+\cdots\;,
\end{equation}
which always terminates for a finite order operator $C_I$. In each term of
such $\bar O$ expansion series, by shifting all destruction operators $C_I$ 
to the right, and using the property $C_I|\Phi\rangle =0$,
we conclude that only terms containing constants or only creation
operators survive. We therefore have a general expression
\begin{equation}
O(C^\dagger_I,C_I)|\Psi_g\rangle = {\cal O}
(C^\dagger_J,F_J)|\Psi_g\rangle\;,
\end{equation}
where ${\cal O}(C^\dagger_J,F_J)$ is a function containing up to linear 
terms in $C^\dagger_J$ and finite-order terms in $F_J$. We shall refer 
Eq.~(19) as \textbf{linear theorem} in our variational approach as
it is useful for general analysis. In fact, the important Eqs.~(5) and (6) 
in Sec.~II are two specific application of this linear theorem. Therefore 
we can write, for a special case of Eq.~(19),
\begin{equation}
C_{L'}C^\dagger_L|\Psi_g\rangle = Y_{L'L}(C^\dagger_I,F_I)|\Psi_g\rangle\;,
\end{equation}
where $Y_{L',L}(C^\dagger_I,F_I)$ is a function containing up to linear
terms in $C^\dagger_I$ and finite-order terms in $F_I$. Using Eq.~(20), 
Eq.~(16) can be written as
\begin{equation}
I_e=I_g\sum_{L,L'}\tilde x_{L'}Y_{L'L}(\tilde g_I,F_I)x_L\;.
\end{equation}
Combining with the optimal condition of Eq.~(15), it is easy to show
\begin{equation}
\frac1I_e \langle\tilde \Psi_g|H\tilde XX|\Psi_g\rangle=E_g\;.
\end{equation}
Hence, we obtain similar equation to Eq.~(14) for the energy difference,
\begin{equation}
\epsilon=\frac{1}{I_e}\langle\tilde\Psi_g|\tilde XHX|\Psi_g\rangle-E_g
 =\frac{I_g}{I_e}\langle\tilde X [H, X]\rangle\;.
\end{equation}

We now apply the above formulas to discuss quasiparticle excitations of
spin systems of Eq.~(1). For simplicity, we consider an approximation in
which we retain only single-spin-flip operators in $X$ and $\tilde X$ of 
Eqs.~(12) and (13), 
\begin{equation}
X\approx \sum_i x_i s^-_i\;,\quad \tilde X\approx\sum_i \tilde x_i s^+_i\;,
\end{equation}
with coefficients chosen as
\begin{equation}
x_i =x_i(q)=\sqrt{\frac2N}e^{i\bf{q\cdot r_i}}\;,\quad
\tilde x_i =\tilde x_i(q)=\sqrt{\frac2N}e^{-i\bf{q\cdot r_i}}\;,
\end{equation}
to define a linear momentum $\bf q$. Such an excited state,
$|\Psi_e\rangle =X|\Psi_g\rangle$, is therefore in the sector of
$s^z_{\rm total}=-1$ and has a linear momentum $\bf q$.
The normalization integral of Eq.~(24) is easily calculated as
\begin{equation}
\frac{I_e}{I_g}=\langle\tilde XX\rangle = 
2\sum_i\tilde x_ix_i\langle s^z_i\rangle +
2s\sum_{i,i',j}\tilde x_{i'}x_if_{i'j}\tilde g_{ij}-
\sum_{i,i',j,j'}\tilde x_{i'}x_if_{i'j}f_{i'j'}
   \tilde g_{ij,i'j'}\;,
\end{equation}
and using Eqs.~(8) and (11), we derive,
\begin{equation}
\frac{I_e}{I_g}=2(s-\rho)(1+\rho_q),
\end{equation}
where $\rho_q \equiv f_q\tilde g_q$. Using approximations 
of Eqs.~(9-11), we obtain, for isotropic point $A=1$,
\begin{equation}
I_e \propto \frac1q\;,\quad q\rightarrow 0\;,
\end{equation}
in all dimensions.

Calculation of the numerator in Eq.~(23) is slightly more complicated.
We quote the result here as, to the order of $(2s)^2$,
\begin{equation}
\langle\tilde X[H,X]\rangle \approx 2s^2z(1+1\rho_q+\gamma_q g_q)\;.
\end{equation}
The energy spectrum of Eq.~(23) is therefore given by, to
the order of $(2s)$,
\begin{equation}
\epsilon_q = \frac{I_g}{I_e}\langle\tilde X[H,X]\rangle
   \approx sz\frac{1+1\rho_q+\gamma_q g_q}{1+\rho_q}\;.
\end{equation}
Using Eqs.~(9-11), we obtain the energy spectrum as
\begin{equation}
\epsilon_q =sz\sqrt{1-(\gamma_q)^2}\;,
\end{equation}
which agrees exactly with the spin-wave theory \cite{and}.
Spectrum of Eq.~(31) is gapless in
any dimension because $\epsilon_q \propto q$ as $q\rightarrow0$.
Similar calculations using spin-flip operators $s^\dagger_j$ and
$s^-_j$ for the $j$-sublattice in Eq.~(24) will produce the same
spectrum as Eq.~(31) except that the corresponding excitation
state has spin $s^z_{\rm total}=+1$. These spin-wave excitations
are often referred to as magnons.

\section{quasiparticle-density-wave excitations}

In previous section, by using quasiparticle
operators (i.e., spin-flip operators $s^\pm$),
we have reproduced the magnon excitations with spin
equal to $+1$ or $-1$. These quasiparticles in general interact with
one another, thus producing quasiparticle density fluctuations.
Excitation states due to these fluctuations are usually best discussed
in terms of corresponding density operator. For our spin
models, density operators are clearly given by operators $s^z$ as it measures
the number of spin-flips with respect to the N\'eel model state and
its expectation value is the order parameter as given by Eq.~(8). For
general purpose, we use notation $C^0_L$ for the quasiparticle
density operators as opposed to the quasiparticle operators
$C_L^\pm$ used earlier. The efficiency of using density operators
to investigate collective modes of a quantum interacting
system was demonstrated by Feynman for the phonon-roton spectrum 
of quantum fluid helium-4 \cite{fey}, who extended Bijl's 
theory \cite{bij} in a much simpler and clearer fashion.
Feynman's excitation formula was also derived by Pines for the
plasmon spectrum of 3D metals \cite{pine}. The 2D plasmon spectrum
first derived by Stern \cite{ster} can also be derived by using
density operator as shown in a PhD thesis \cite{yx_thesis}.
It is interesting to note that both the CBF method for the
ground state and Feynman's theory for excitation states have
been successfully applied to fractional quantum Hall effects \cite{lau,gir}.
Feynman's excitation theory is now often referred to as single-mode
approximation \cite{gir}.

Following Feynman, we write the quasiparticle density-wave excitation
state as
\begin{equation}
|\Psi_e^0\rangle = X^0|\Psi_g\rangle, \quad
X^0= \sum_L x_L C^0_L,
\end{equation}
where, as defined earlier, $C^0_L$ are the quasiparticle density
operators. The bra state is given by the hermitian conjugate of Eq.~(32),
$\langle\tilde\Psi_e|=\langle\tilde\Psi_g|\tilde X^0$.
Using the same argument as before for the quasiparticle excitation of
Eq.~(23), we obtain a similar equation for the energy difference
for our collective modes as,
\begin{equation}
\epsilon^0=\frac{I_g}{I^0_e}\langle\tilde X^0 [H, X^0]\rangle\;,
\end{equation}
where $I^0_e=\langle\tilde\Psi_e^0|\Psi^0_e\rangle$. We notice that, by definition, 
density operator $C^0_L$ is a hermitian operator, $(C^0_L)^\dagger=C^0_L$.
By considering a similar excited state $\tilde X^0|\Psi_g\rangle$, it is 
straightforward to derive the following double commutation formula,
\begin{equation}
\epsilon^0=\frac{I_g}{2I^0_e}\langle[\tilde X^0, [H, X^0]]\rangle\;.
\end{equation}
The double commutation in the above equation is the key to
the efficiency of Feynman's excitation theory. It is often referred to
as $f$-sum rule in other quantum systems
such as electron gases \cite{pine}.

Before we apply Eq.~(34) for collective modes in spin lattices, it is 
useful to discuss sum rules in our spin models as density operators normally obey
sum rule equations \cite{fee,ripk}. The order parameter of Eq.~(8) can also be 
calculated through two-body functions as
\begin{equation}
(\langle s^z_i\rangle)^2=\frac{\langle\tilde\Psi_g|(s^z_a)^2|\Psi_g\rangle}
{\langle\tilde\Psi_g|\Psi_g\rangle},
\end{equation}
where $s^z_a= \sum_l(-1)^l s^z_l/N$ is the staggered spin operator.
We introduce total magnon-density operator $\hat n_i$ as
\begin{equation}
2\hat n_i = 2s-s^z_i+\frac1z\sum_{n=1}^z s^z_{i+n},
\end{equation}
where as before summation over $n$ is over all $z$ nearest neighbors.
Hence, the sum rule for the one-body function is simply
$\frac2N\sum_i\langle \hat n_i\rangle=\rho$. The two-body
Eq.~(35) can now be written as, using translational invariant 
property $\rho_i=\rho$,
\begin{equation}
\frac2N\sum_{i'=1}^{N/2}\langle \hat n_i\hat n_{i'}\rangle=\rho\rho_i
 = \rho^2,
\end{equation} 
which is the familiar two-body sum rule equation \cite{fee,ripk}. 
In the approximation of 
Eq.~(7-11), we find that this sum rule is obeyed in both cubic and square 
lattices in the limit $N\rightarrow\infty$. In particular, we find that
$(\frac2N\sum_{i'}\langle \hat n_i\hat n_{i'}\rangle - \rho^2)
\propto 1/N$ in a cubic lattice and $\propto (\ln N)/N$ in a square 
lattice. These asymptotic properties are important in the
corresponding excitation states as will eb
discussed later. However, Eq.~(37) is violated in the one-dimensional
model, showing the deficiency of the two-spin-flip approximation of Eq.~(7)
for the one-dimensional model. We therefore leave further investigation
elsewhere and focus on the cubic and square
lattices in the followings, using approximations of Eqs.~(7-11).

We therefore write our magnon-density-wave excitation state using 
total magnon density operator $\hat n_i$ of Eq.~(36) as,
\begin{equation}
|\Psi_e^0\rangle = X^0_q|\Psi_g\rangle, \quad
X^0_q = \sum_i x_i(q) \hat n_i, \quad q>0
\end{equation}
and its hermitian counterparts for the bra state,
$\langle\tilde \Psi_e^0| = \langle\tilde\Psi_g|\tilde X^0_q$. The
coefficient, $x_i(q)=\sqrt{\frac2N}e^{i\bf{q\cdot r_i}}$, etc. 
The condition $q>0$ in Eq.~(38) ensures the orthogonality between this
excited state with the ground state. The excitation energy difference
is given by Eq.~(34) as
\begin{equation}
\epsilon^0_q = \frac{N(q)}{S^0(q)},\quad q>0
\end{equation}
where $N(q)\equiv\langle[\tilde X^0_q,[H,X^0_q]]\rangle/2$ and
$S^0(q)\equiv\langle\tilde X^0_qX^0_q\rangle$ is the structure
function. Both $N(q)$ and $S^0(q)$ can be
straightforwardly calculated as, using approximations of Eqs.~(7-11),
\begin{equation}
N(q) =-\frac{sz}{2}\sum_{q'}
(\gamma_{q'}+\gamma_q\gamma_{q-q'})\tilde g_{q'},
\end{equation}
and
\begin{equation}
S^0(q) = \frac14(1+\gamma^2_q)\rho + \frac14 \sum_{q'}[(1+
 \gamma^2_q)\rho_{q'}\rho_{q-q'}+
 2\gamma_q \tilde g_{q'}\tilde g_{q-q'}]\;,
\end{equation}
where $q>0$. The energy spectrum $\epsilon^0_q$ of Eq.~(39) can then
be calculated numerically. We notice that Eq.~(41) is closely related to 
the the sum rule Eq.~(37) which correspond to $q=0$ case [with an
additional term in Eq.~(41) when $q\rightarrow0$]. Using the approximation
of Eqs.~(9) and (10), it is not difficult to show that $N(q)$ of Eq.~(40) has a
nonzero, finite value for all values of $q$. Any special feature
such as gapless in the spectrum $\epsilon^0_q$ therefore comes from the
structure function of Eq.~(41), and hence is determined by 
the asymptotic behaviors of the sum rule Eq.~(37) mentioned earlier.

For a cubic lattice, we plot $N(q)$ and $S^0(q)$ for two regions
of ${\bf q}$ in Fig.~1.  In Fig.~2 we plot the corresponding
spectra of Eq.~(39), together with that of magnon excitations of Eq.~(31)
for comparison. As can be seen from Fig.~2, the spectrum $\epsilon^0_q$ has
a nonzero gap everywhere. The minimum gap is about $\epsilon^0_q\approx 0.96sz$
at ${\bf q}=(q_0,q_0,q_0)$ with $q_0\approx0.04\pi$. (This is slightly different
to that reported in Ref.~12 where detailed calculations in this region
had no been were done.) This gap is about the same as the largest magnon 
energy, $\epsilon_q = sz$ at ${\bf q}=(\pi/2,\pi/2,\pi/2)$
from Eq.~(31). At ${\bf q}=(\pi/2,\pi/2,\pi/2)$,
we have the largest energy $\epsilon^0_q \approx 2.92sz$. This
is nearly three magnons' energy at this $\bf q$. At ${\bf q}=(\pi,0,0)$,
we obtain $\epsilon^0_q\approx2.56sz$. 

For a square lattice, the structure function $S^0(q)$ of Eq.~(41)
has a logarithmic behavior $\ln q$ as $q\rightarrow0$. This is not surprising as
discussed earlier in the sum rule Eq.~(37), where occurs the asymptotic behavior
$(\ln N)/N$ as $N\rightarrow\infty$. For small values of $q$, $N(q)$ approaches
to a finite value, $N(q)\approx 0.275sz$ as $q\rightarrow0$. The corresponding energy
spectrum of Eq.~(39) is therefore gapless as $q\rightarrow 0$. 
Similar to the cubic lattice, we plot $N(q)$ and $S^0(q)$ of a square
lattice in Fig.~3 and the corresponding spectra of Eq.~(39) and Eq.~(31) 
in Fig.~4. As can be seen from Fig.~4, magnon-density-wave energy 
is always larger than the corresponding
magnon energy. At small values of $q$ ($q<0.05\pi$), we find a good 
approximation by numerical calculations for the structure function,
$S^0(q)\approx 0.31-0.16\ln q$ with $q_x=q_y$. Similar behavior holds
near ${\bf q}=(\pi,\pi)$. The energy spectrum of Eq.~(39) in these region can
therefore be approximated by
\begin{equation}
\epsilon^0_q \approx \frac{0.275sz}{0.31-0.16\ln q},\quad q\rightarrow0
\end{equation}
for a square lattice with $q_x=q_y$. We notice the slight difference
for the coefficients of Eq.~(42) to that of Eq.~(19) of Ref.~12 where we
focused in the region with $q_y=0$. Although our calcuations clearly show 
this spectrum of a square lattice is gapless at $q=0$ 
and ${\bf q}=(\pi,\pi)$, it is nevertheless very "hard" when comparing
with the magnon's soft mode $\epsilon_q\propto q$ at small $q$. For example,
we consider a system with lattice size of $N = 10^{10}$,
the smallest value for $q$ is about $q\approx 10^{-10}\pi$ and we have
energy $\epsilon^0_q \approx 0.07sz$. Comparing this value with the 
corresponding magnon energy $\epsilon_q\approx 10^{-10}sz$, we conclude
that the energy spectrum of Eq.~(39) is "nearly gapped" in a square lattice.
We also notice that the largest energy in a square lattice 
$\epsilon^0_q\approx 2.79sz$ at ${\bf q}=(\pi,\;0)$, not at 
${\bf q}=(\pi/2,\;\pi/2)$ as the case in a cubic lattice.
At ${\bf q}=(\pi/2,\;\pi/2)$, we obtain $\epsilon^0_q\approx 2.62sz$ for
the square lattice. We will discuss physical implications of these
excitations in the next section.

\section{Discussion}

We have obtained in this article two main results. Firstly, we have
succeeded in extending our recently proposed variational approach 
to describe, in general terms,
excitation states of a quantum many-body system. Secondly, we have applied 
our technique to quantum antiferromagnets thus reproducing the 
well-known magnon excitations and, in addition, we have obtained
a new, spin-zero longitudinal collective modes which have been missing
in spin-wave theory of Anderson \cite{and}. In the followings, we
shall discuss further physical implications of these new 
excitations and we conclude this article with a summary in the end.

It is interesting to notice similar behaviors between collective modes
of quantum antiferromagnets and plasmon excitations of electron gases as both
spectra show a large energy gap in 3D and are gapless in 2D. In fact, 
further similarity between these two quantum systems can be made.
It is generally accepted that, for many purposes, a quantum antiferromagnet
at zero temperature can be considered as a gas of weakly interacting,
equal numbers of spin $\pm1$ magnons (the transverse 
spin-flip wave excitations with respect to the classical N\'eel state); also
present in the system are the spin-zero, longitudinal fluctuations consisting
of multi-magnon continuum \cite{brau,bunk,schw,hube}. This is similar
to quantum electron gases which can also be considered as a gas of 
weakly interacting, equal numbers of quasielectrons and 
holes (the transverse excitations near the Fermi surfaces) and the charge-neutral,
longitudinal fluctuations producing quasielectron-hole continuum \cite{pine}.
Plasmon excitation of electron gases have been well observed as
sharp peaks over the electron-hole continuum \cite{pine}. However, plasmon-like
collective modes of quantum antiferromagnets as discussed in this article
have so far eluded from observation to our best knowledge. We can only draw 
some support by considering a finite-size Heisenberg model of Eq.~(1).
As the ground state of a finite antiferromagnetic Heisenberg lattice
is spin-singlet, we expects lowing-lying excitations are triplet
with $z$-component of spin equal to $0,\pm1$. As lattice size
increases from finite to infinite, for the cubic and square lattices, 
the spontaneous-symmetry-breaking occurs and the ground state is no longer 
spin singlet but has a long-ranged antiferromagnetic order. We expect
that the triplet excitation splits into different branches. The magnon 
spectrum of Eq.~(31) with spin $\pm1$ and the spectrum of Eq.~(39) for spin-zero
magnon-density waves are our approximation for these different branches of 
excitations. We also notice that recently modified spin-wave theories 
were applied to finite systems with results in reasonable agreements with 
exact finite-size calculations \cite{aa,ht,mt}. As pointed out in Ref.~28, 
however, a major deficiency in this theory is the missing 
spin-zero excitations as the low-lying excitations for a finite lattice 
Heisenberg model are always triplet as mentioned earlier. We believe
our magnon-density-wave excitation as discussed here corresponds to
the missing branch; the energy gap in the cubic lattice and the
nearly gapped spectrum in the square lattice of Eq.~(39) reflect the
nature of long-ranged N\'eel order in the ground states of infinite
systems. Improvement for spectra of Eq.~(39) can be done in similar
fashion as was done for the ground state detailed in paper II, particularly
for the square lattice. We will
have more motivation to do so if we have experimental evidence of
these collective modes.

In any case, this artcle concludes our general presentation of
a new formalism of variational coupled-cluster method
for a quantum many-body system. Beginning in paper I, we
introduced and discussed bare distribution functions, key ingredient of
this formalism. In paper II, we developed diagrammatic
techniques for practical, high-order calculations of these
functions. Application to quantum antiferromagnets has demonstrated
the efficacy of this technique. Present artcle extends
this formalism to excitation states. As discussed earlier,
application to quantum antiferromagnets have produced new modes
which have been missing in all spin-wave theories and are
yet to be confirmed by experiment. Our next main focus is to combine
our present variational approach with the CBF method as first discussed
in paper II. Hence we write our new ground state as
\begin{equation}
|\Psi_u\rangle =e^{S^0}|\Psi_g\rangle = e^{S^0}e^S|\Phi\rangle\;,
\end{equation}
where $S$ is as given by Eqs.~(2) and $S^0$ is the generalized
Jastrow correlation operator involving quasiparticle density operators as
\begin{equation}
S^0 = \sum_{ij}f^0_{ij}s^z_is^z_j\;,
\end{equation}
with $f^0_{ij}$ as new variational functions. Using the 2-spin-flip approximation
of Eq.~(7) for $S$, the new wave function of Eq.~(46) can be understood as including
both quasipartcle fluctuations described by operator $\exp(S)$ and 
quasiparticle-density fluctuations described by operator $\exp(S^0)$. The results of
collective modes obtained in Sec.~IV certainly make this combination of 
Eq.~(43) much more appealing and imperative.

\acknowledgments I am grateful to members of our Department,
particularly to members of former UMIST Physics Department before the
merger, for their continuous support. I am also grateful to C.E. Campbell
for useful discussions on the CBF method, which have inspired parts of 
this work.

\newpage

\begin{figure}

Fig.~1 $N(q)$ and $S^0(q)$ of Eqs.~(40) and (41) for a cubic lattice. Shown are
the values for two regions ${\bf q}=(0,0,0)$ to $(\pi, 0,0)$ and to 
$(\pi, \pi,\pi)$.

Fig.~2 Excitation energy spectra in unit of $sz$ in 
a cubic lattice. The higher branch is for the 
plasmon-like excitation of Eq.~(39) and the lower one is for the magnon 
excitation of Eq.~(31). 

Fig.~3 Similar to Fig.~1 but for a square lattice. The divergence of $S^0(q)$
at ${\bf q}=(0,0)$ and $(\pi,\pi)$ is given in the text.

Fig.~4 Similar to Fig.~2 but for a square lattice. The behavior near ${\bf q}
=(0,0)$ and $(\pi,\pi)$ for magnon-density-waves is given by Eq.~(42).

\end{figure}

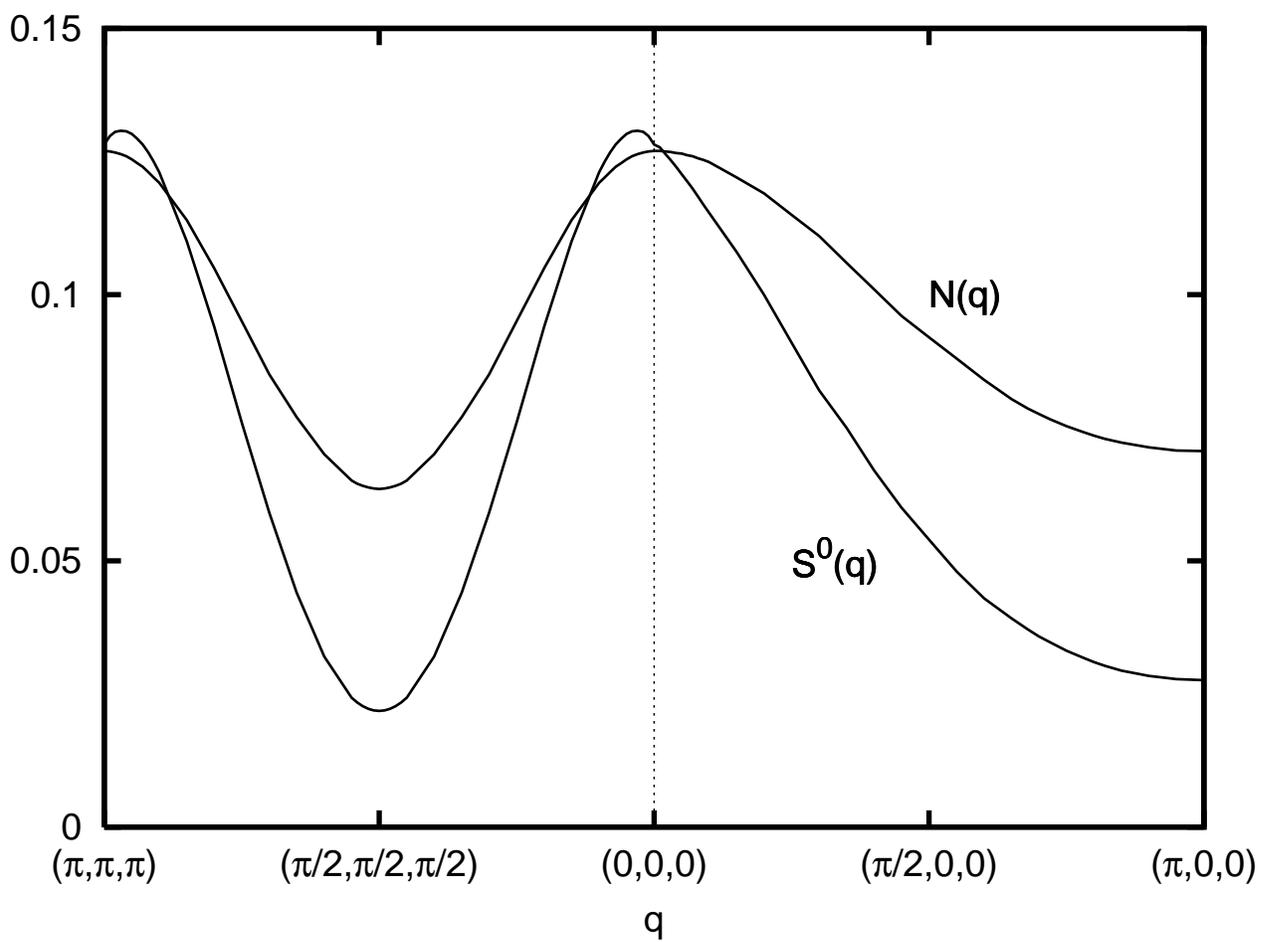
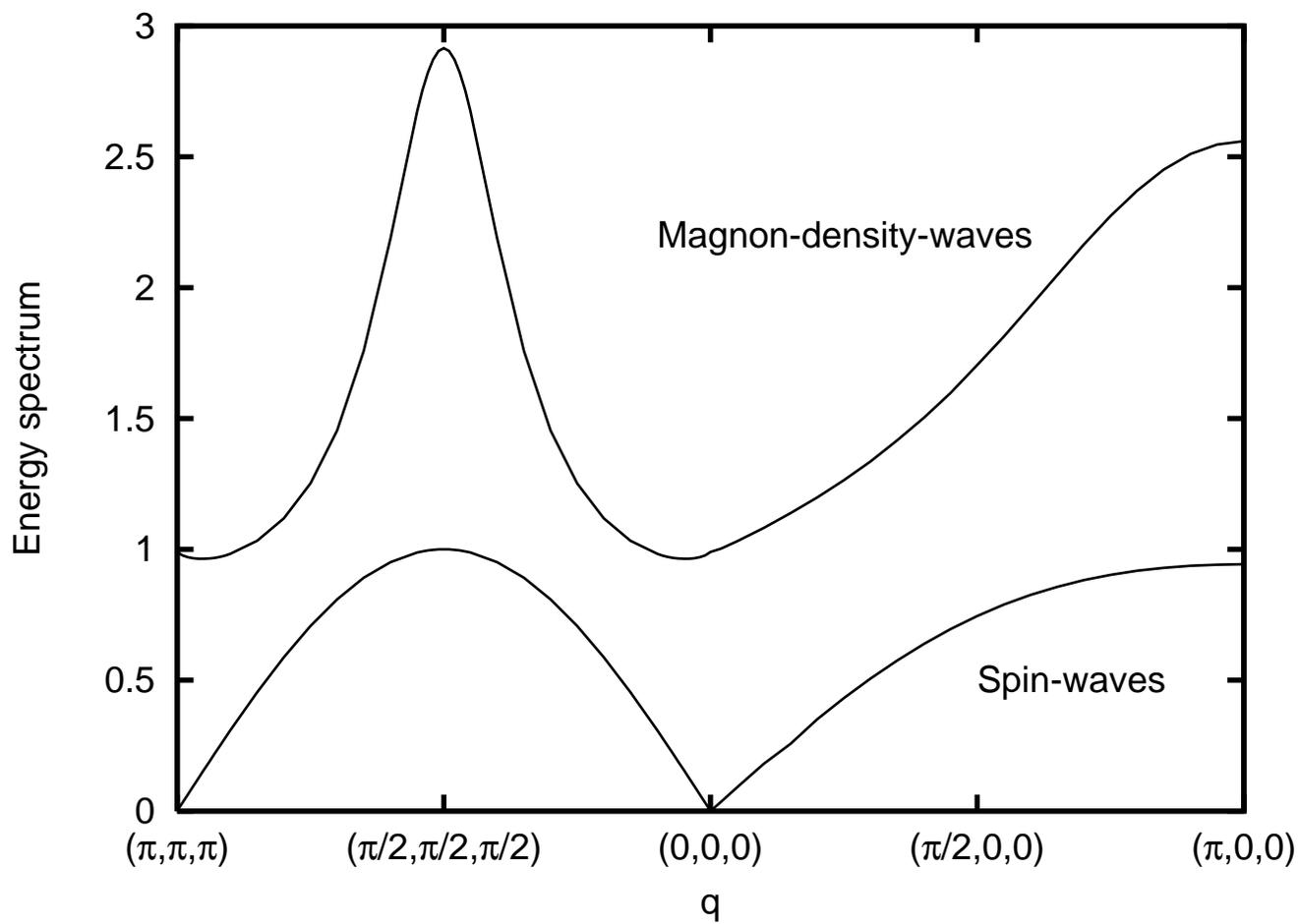
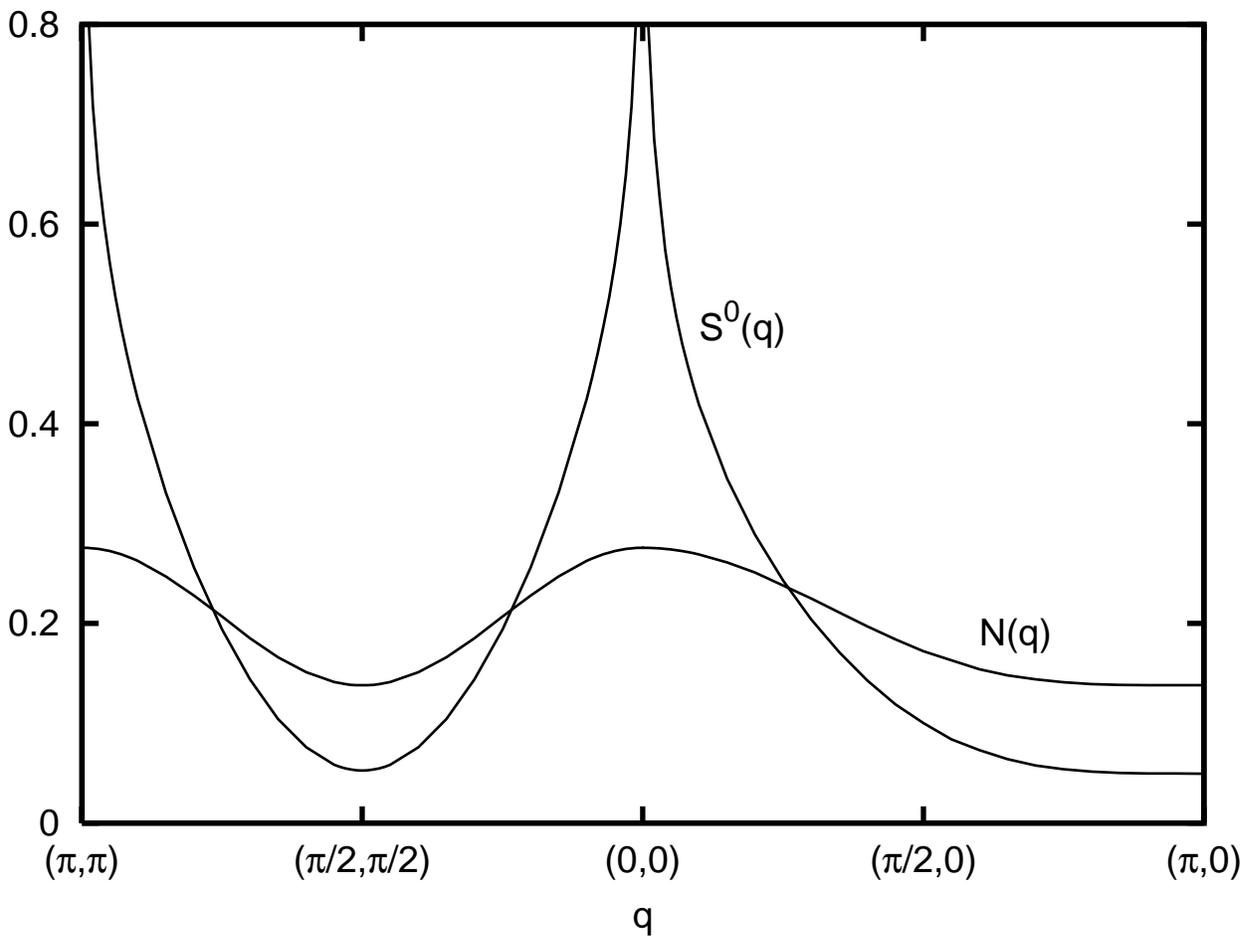
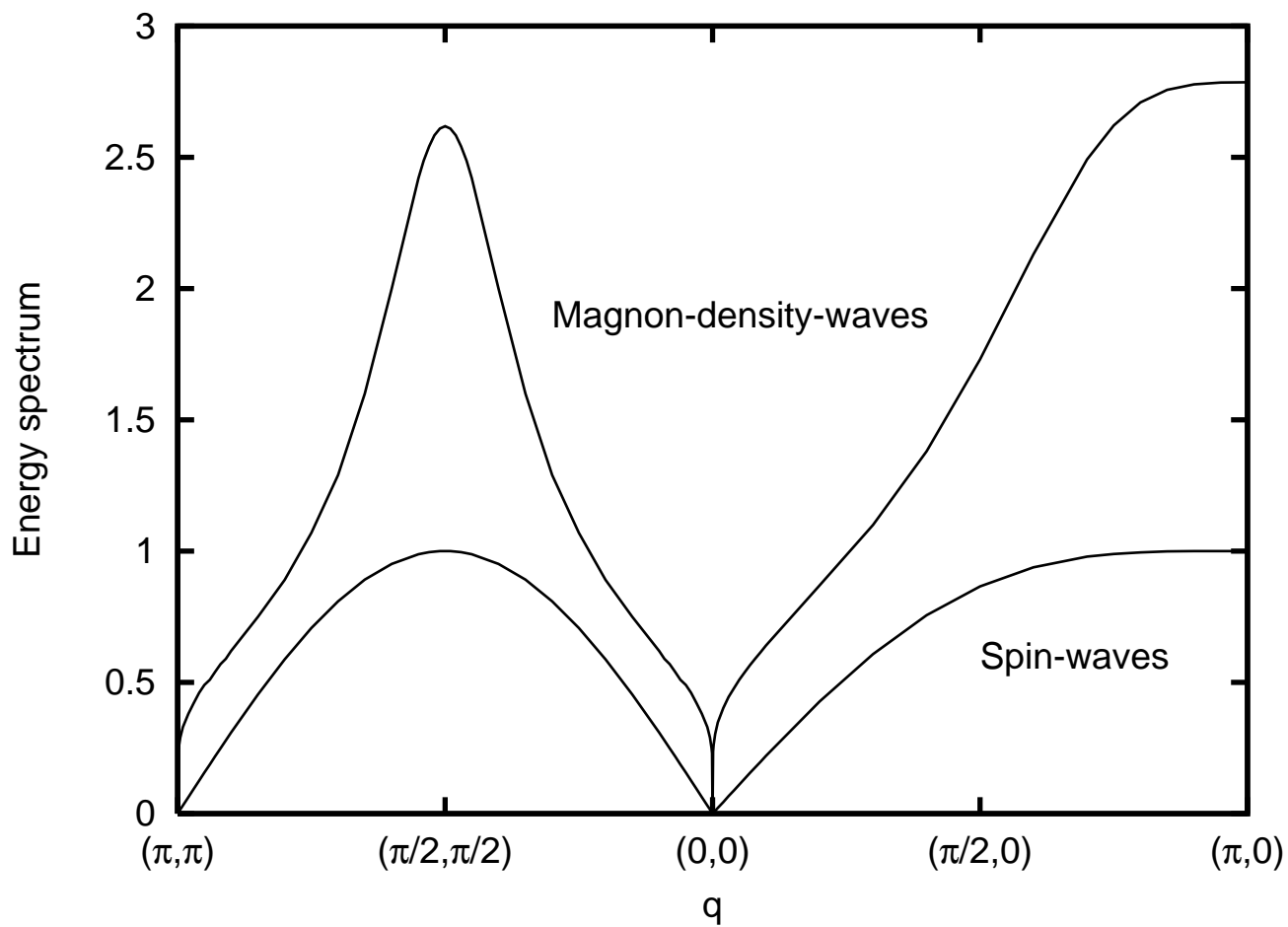
\end{document}